# The Fabric of Space-time

**Charles Francis**

**Abstract:**

The formulae of special relativity are developed through the *k*-calculus with no presumption of a manifold. The metric is determined empirically by the exchange of photons, and the treatment suggests that the exchange of photons seen in quantum electrodynamics is also responsible for the "fabric" of Minkowsky space-time. This suggests that on quantum scales the point-like nature of elementary particles, should appear "defocussed" or "fuzzy". Further work on the implications in the laws of physics is referred to.

Charles Francis
Clef Digital Systems Ltd
Lluest, Neuaddlwyd
Lampeter
Ceredigion
SA48 7RG

22/6/99

# The Fabric of Space-time

## 1    Introduction

There is increasing interest in the idea that fundamental variables such as time should be discrete [1], and there are many references in the literature on the potential quantisation of gravity which suggest that at a fundamental level space-time may be discrete [2]. This paper is one of a series examining the consequences for the laws of physics of replacing the assumption of a pre-existent space-time continuum with the observation that time and distance are numbers produced by a measuring apparatus.

The purpose of the present paper is to show that the laws of special relativity can derived from an empirical treatment of measurement which does not require the assumption of a manifold. The import of the treatment is to demonstrate that Minkowsky space-time is a mathematical construction having no direct ontological analogue. Since the assumption of empty space is not required to derive the mathematical laws of special relativity, it has no impact on the theory. The introduction of ontological space-time would either give wrong results or make no change to the empirical predictions of the theory, and so be untestable and scientifically meaningless. Instead we describe the manner in which combinations of particle interactions may generate the mathematical properties of Minkowsky space-time.

An interesting corollary to this idea is that not only motion should be regarded as relative in the special theory of relativity [3]. The position of a body or particle is also defined relative to other matter. It is intuitively clear that, in the absence of a pre-existent manifold, uncertainty will be a feature of the empirical definition of co-ordinate systems, and that this uncertainty is not classical probability theory of an unknown variable, and will require an alternative mathematical treatment. Further work is referred to suggesting that relativity in position is the principle underlying uncertainty in the laws of quantum mechanics as well as other fundamental laws in physics.

## 2    Co-ordinate Systems

There is room for confusion between two very similar questions, 'What is time?' and 'What is the time?'. The first question has something to do with consciousness, and our perception of time as a flow from past to future. It admits no easy answer, but is quite distinct from the second question and only the second question is relevant in the definition of space-time co-ordinates. The answer to the question 'What is the time?' is always something like 4:30 or 6:25.

**Definition:** The time is a number read from a clock.

There are many different types of clock, but every clock has two common elements, a repeating process and a counter. The rest of the mechanism converts the number of repetitions to conventional units of time. A good clock should provide accurate measurement and it should give a uniform measure of time. We cannot count less than one repetition of the process in the clock, so for accurate measurement the process must repeat as rapidly as possible. In a uniform clock, the repeating process must repeat each time identical to the last, uninfluenced by external matter. One repetition gives the minimum unit of time for any given clock. Subdividing this unit of time requires a second clock. So time takes integer values. In principle there may be clocks, i.e. repeating processes, which are faster than any process used in a practical clock, but there must be some indivisible process, which determines a smallest notional unit of time, the chronon, called after its name in antiquity. There may be more than one such indivisible repeating process, so the chronon need not be unique. I assume that is very much smaller than the unit given by any practical clock, so that for practical purposes conventional measures of time can be regarded as (large) whole numbers of chronons.

A clock defines the time, but does so only at one place. A space-time co-ordinate system also requires a definition of distance, and a definition of time at a distance from the clock. This is provided for by the radar method, but in practice the frequency of the electromagnetic radiation is irrelevant, and the definition refers to light of any frequency.

**Definition:** The distance of an event is half the lapsed time for light to go from the clock to the event and return to the clock. the time at which the signal is reflected is the mean time between when it is sent and when it returns.

**Reason:** As Bondi pointed out *"with our modern outlook and modern technology the Michelson-Morley experiment is a mere tautology"* [5]. It tautologically defines space-time co-ordinates only at points where the radar method is actually used.

The radar method defines distance in units of time, so space-time co-ordinates are strictly elements of $\mathbb{N}^4$. Radar is preferred to a ruler, because it applies directly to both large and small distances, and because a single measurement can be used for both time and space co-ordinate. The radar method also measures direction and it will be seen that the algebra is formally identical for 3-vectors with a Euclidean metric and for one dimensional space-time diagrams, such as figure . Each point on a space-time diagram represents an event.

Space-time diagrams are defined such that lines of equal time are horizontal and lines of equal distance are vertical (figure 1). By definition, uniform motion in the reference frame is shown by a straight line on the diagram. To use radar we must know the speed of light (if distance were defined using a ruler, then to measure the time at an event we would still need to know the speed of a message from the event). But now we have a paradox. To measure speed we conduct a time trial over a measured distance, but first time must be defined at both ends of the ruler, which requires knowledge of the speed of light. We know no other way to measure the time of an event at a distance from a clock; if we synchronise two clocks by bringing them together, we have no guarantee that they remain synchronised when they are separated, unless light is used to test their synchronisation. Thus the speed of light is an absolute constant because measurement of speed requires a co-ordinate system, which requires light for its definition. An experiment to determine the speed of light actually measures the conversion factor from natural units in which the speed of light is 1. Thus, by definition, light is drawn at $45^\circ$.

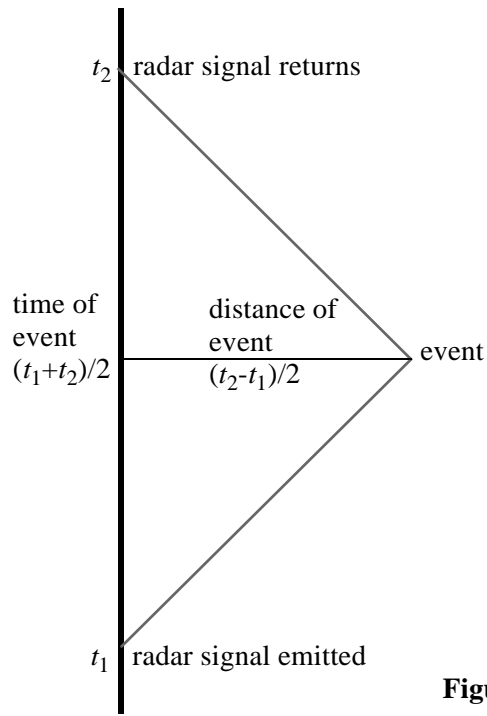

**Figure 1**

**Definition:** A space time co-ordinate system defined by radar is known as a reference frame.

By this definition, a reference frame is a mathematical construction, namely the set of all values which can result from process of measurement, not a physical entity. It depends on the possibility of measurement and cannot be extended indefinitely into space or defined in a perfect vacuum where there is no matter.

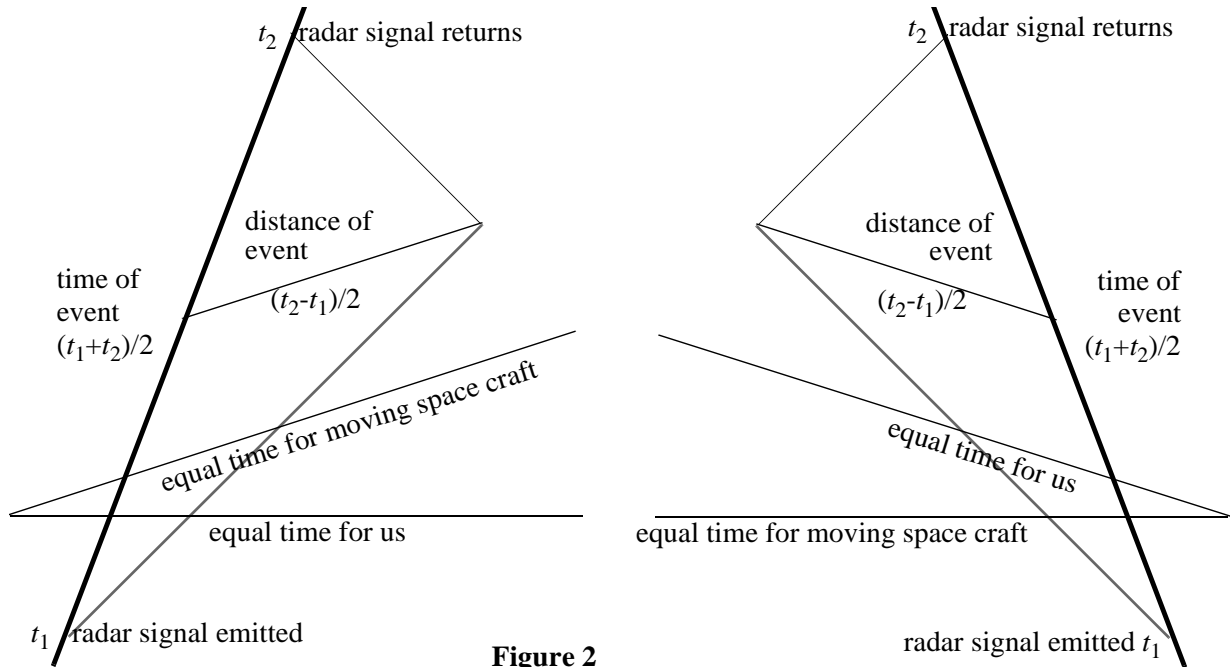

**Figure 2**

Once clocks are separated, there is no way to synchronise them directly, but, according the principle of homogeneity, two clocks will give the same unit of time if the physical processes in each are identical. If we wish to compare our coordinate system with the coordinate system of a moving observer, we need to know what unit of time the moving observer is using. Once clocks are separated, there is no way to synchronise them directly, but, according the principle of homogeneity, two clocks will give the same unit of time if the physical processes in each are identical. Figure 2 shows the coordinate system defined by an observer in a moving space craft, as it appears to us, and our coordinate system as it appears to him. The moving observer represents himself with a vertical axis, and he would draw us at an angle. In his diagram our reference frame appears distorted. .

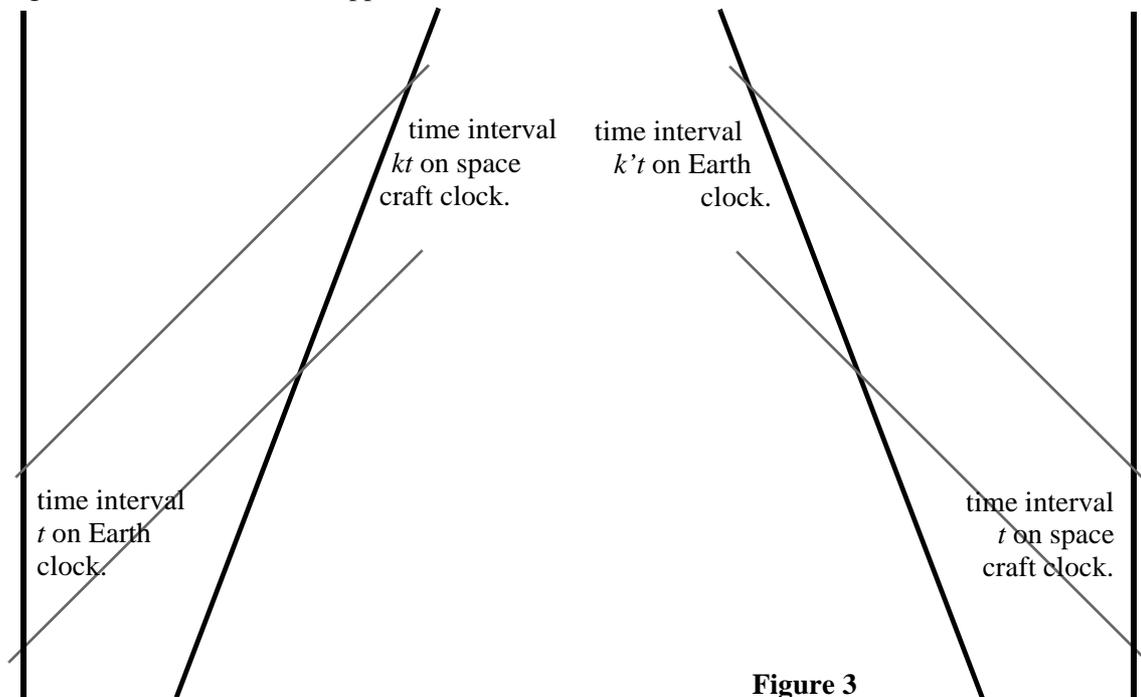

**Figure 3**

In figure 3, a space craft is uniformly moving in the Earth's reference frame. The space craft and the Earth have identical clocks and communicate with each other by radio or light. The Earth sends the space craft two signals at an interval *t*. The space craft receives them at an interval *kt* on the space craft's clock. $k \in \mathbb{R}$ is immediately recognisable as red shift. Although *kt* is not necessarily an integer, its fractional part is less than a chronon, and is lost in measurement. Similarly if the observer on the space craft sends two signals at an interval *t* on his clock, they are received at an interval *k't* on the Earth.

There is no fundamental difference between the matter in the space craft and the matter in the Earth. The space craft can be regarded as stationary, and the Earth as moving. The principle of homogeneity implies that signals sent by the space craft to the Earth are also subject to red shift. The defining condition for the special theory of relativity is that there is a special class of reference frames such that

**Definition:** For inertial reference frames red shift is both constant and equal for both observers, $k = k'$.

**Definition:** The law of co-ordinate transformation between inertial reference frames is Lorentz transformation.

We know from observation that inertial reference frames exist, at least to the accuracy of measurement and they will be assumed in this paper. The general theory of relativity places a more general condition on red shift. The implication will be studied in another paper, currently in draft, in which it is shown that, as a direct result of the discrete nature of particle interactions, the inherent delay in the return of the signal forces the use of non-inertial frames (such that $k \neq k'$) and results in the force of gravity.

**Theorem:** (Time dilation, figure 4) The time *T* measured by a space craft's clock during an interval *t* on the Earths clock is given by

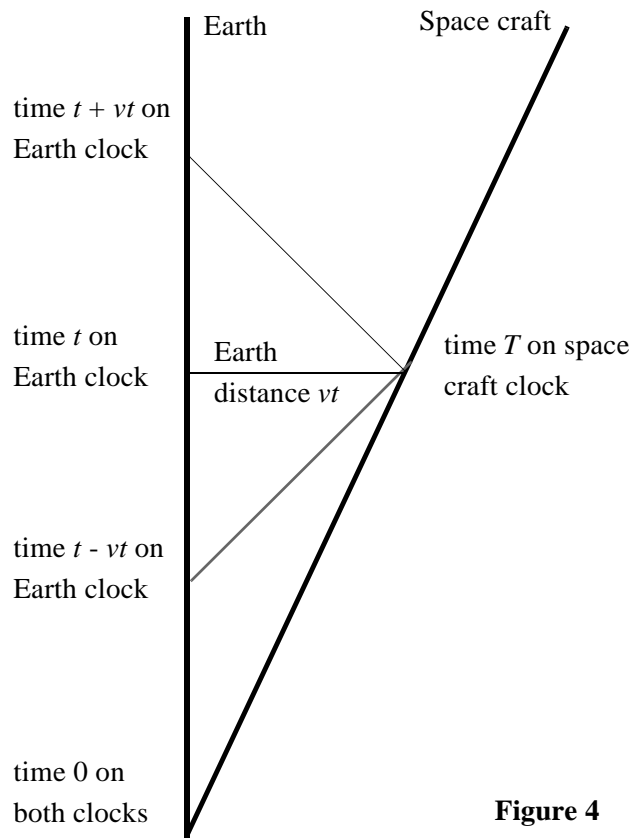

2.1 $\quad T = t\sqrt{1 - v^2}$

**Proof:** The space craft and the Earth set both clocks to zero at the moment the space craft passes the Earth. The space craft is moving at speed *v*, so by definition, after time *t* on the Earth clock, the space craft has travelled distance *vt*. Therefore Earth's signal was sent at time *t - vt,* and returned at time *t + vt*. For inertial reference frames, if the space craft sends the Earth signals at an interval *t* the Earth receives them at an interval *kt*., so

2.2 $\quad T = k(t - vt)$.

Then by applying the Doppler shift again for the signal coming back

2.3 $\quad t + vt = k^2(t - vt)$

Eliminating *k* gives 2.1, the formula for time dilation.

**Figure 4**

**Theorem:** (Lorentz Contraction, figure 5)
A distance $d$ on the earth is measured on a space craft to be

2.4 $\qquad D = \dfrac{d}{\sqrt{1-v^2}}$

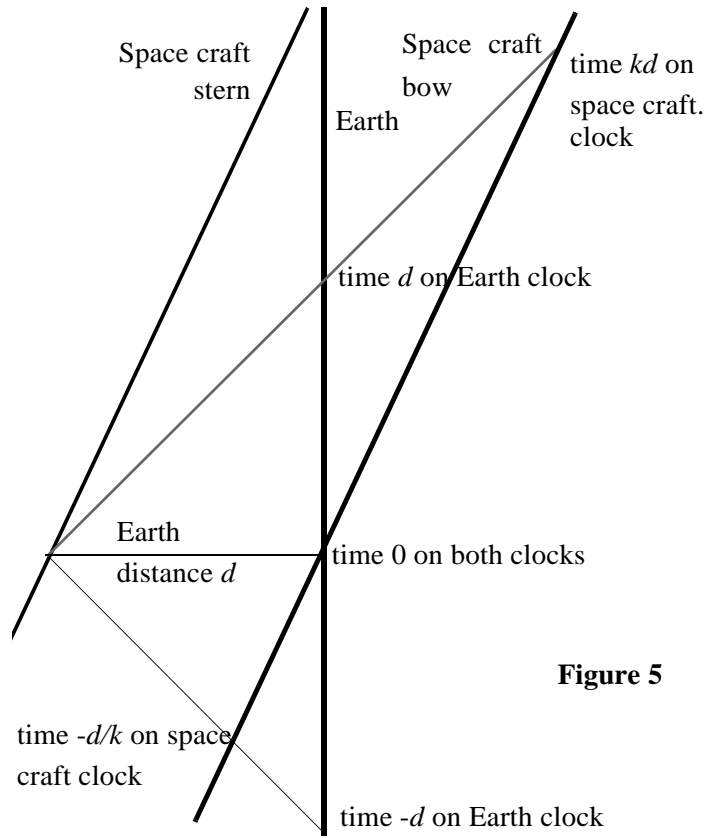

**Figure 5**

**Proof:** The bow and stern of the space craft are shown as parallel lines. The space craft's clock is in the bow. For ease of calculation, both the space craft and the Earth set their clocks to zero when the bow passes the Earth clock. Earth uses radar to measure the distance, $d$, to the stern at time 0. To do so, the signal must have been sent at time $-d$, and must return at time $d$ on the Earth clock. From the Doppler shift, on the space craft's clock, the signal passes the bow of the space craft at time $-d/k$ and comes back at time $dk$. So, according to the moving space craft

2.5 $\qquad D = (dk + d/k)/2$

Eliminating $k$ by 2.3 gives 2.4, the formula for Lorentz contraction.

Laws which are the same in all co-ordinate systems are expressed in terms of invariants, mathematical quantities which are the same in all co-ordinate systems. The simplest invariant is an ordinary number or scalar. Another invariant, familiar from classical mechanics, is the vector. Changing the co-ordinate system has no effect on a vector, but it changes the description of a vector in a co-ordinate system.

**Definition:** A space-time vector is the difference in the co-ordinates of two events. When no ambiguity arises space-time vectors are simply called vectors.

**Theorem:** The mass shell condition

2.6 $\qquad m^2 = E^2 - \boldsymbol{p}^2$

**Proof:** A vector can be represented as a straight line on a space-time diagram, and described by components

2.7 $\qquad r = (E, \boldsymbol{p})$

For a time-like vector, $r$, there is a particular reference frame in which it represents a state of rest, namely when it is aligns with the axis representing the clock on which the definition of that reference frame is based. In this reference frame $r$ has co-ordinates

2.8 $\qquad r = (m, 0)$

An observer moving at velocity $\boldsymbol{v}$ relative to the clock describes $r$ by co-ordinates given by the formulae for time dilation, 2.1 and Fitzgerald contraction, 2.4

2.9 $\qquad r = (E, \boldsymbol{p}) = \left(\dfrac{m}{\sqrt{1-v^2}}, \dfrac{m\boldsymbol{v}}{\sqrt{1-v^2}}\right)$

The mass shell condition, 2.6, follows at once

**Definition:** If $x = (x_0, x_1, x_2, x_3)$ and $y = (y_0, y_1, y_2, y_3)$ are vectors in space-time then the scalar product is

2.10 $\qquad x \cdot y = -x_0 y_0 + x_1 y_1 + x_2 y_2 + x_3 y_3$

**Theorem:** The scalar product is invariant under Lorentz transformation
**Proof:** Straightforward algebra from 2.9

## 3  The Fabric of Space-time

The microscopic structure of Euclidean geometry is described in analysis by use of limits. It has been known from the time of Zeno that the continuous structure of points comprising space cannot be established, and although the notion of absolute space-time has entered into the scientific model, it does not appear that Newton intended that absolute space should be regarded as a perfect description of reality. As he says when introducing the concept in the preface to the *Principia* [6]:

The description of right lines and circles, upon which geometry is founded, belongs to mechanics. Geometry does not teach us to draw these lines, but requires them to be drawn.

Clearly Newtonian space is only postulated to the accuracy possible in engineering. Clifford [7] discussing Riemann's achievement in non-Euclidean geometry also showed awareness that a spacial continuum is an unwarranted assumption:

Riemann has shown us that there are different kinds of lines and surfaces, so there are different kinds of space of three dimensions; and that we can only find out by experience to which of these kinds the piece in which we live belongs. In particular, the axioms of plane geometry are true within the limits of experiment on the surface of a sheet of paper, and yet we know that the sheet is really covered with a number of small ridges and furrows, upon which (the total curvature being not zero) these axioms are not true. Similarly, he says, although axioms of solid geometry are true within the limits of experiment for finite portions of our space, yet we have no reason to conclude that they are true for very small portions; and if any help can be got thereby for the explanation of physical phenomena, we may have reason to conclude that they are not true for very small portions of space.

and Riemann himself says

Either therefore the reality which underlies space must form a discrete manifold, or we must seek the ground of its metric relations outside it. (*translation by Clifford*)

Indeed, Riemann's mathematical definition of a manifold [8] deliberately ignores the question of whether the manifold represents something ontological. As Riemann suggested possible, the treatment of special relativity given here provides for metric relations outside of the assumption of a manifold.

Just as we recognise that the surface of a sheet of paper is fibrous and non-geometrical, we may also conceive that ontological space is not smooth or continuous. While Clifford, *On the Space Theory of Matter*, attempted to describe matter as some sort of twist in the geometry of space, I seek now to describe a "matter theory of space", in which particles are a part of the substructure, or pre-geometry, of space, and space-time is conceived as a fabric of particles. The motivation for this is that quantum electrodynamics has shown that the process used here to define co-ordinate systems, namely the exchange of photons, is also responsible for the electromagnetic force, and so for all the structures of matter in our macroscopic environment. It is not unreasonable, therefore, to postulate that the exchange of photons generates all the geometrical relationships in the macroscopic environment, just as it generates these relationships in the results of measurement by means of radar.

To imagine the substructure of space-time we conceive that each charged particle follows some repeating process according to which we may regard a primitive notion of time as one of its ontological properties. We call this notion a time line. For example a particle may have the possibility of emitting or absorbing a photon in each discrete instant of its time line. This can be considered a repeating process adequate for the notion of primitive time. Whenever an exchange of photons takes place, i.e. a photon is emitted by one charged particle and absorbed by another, and a photon is then immediately emitted by the second particle and absorbed by the first, then a coordinate for the second particle is established in terms of the time line of the first.

Since visualisation involves the awareness of geometry, it is clear that the pre-geometric properties of matter cannot strictly be visualised. Nonetheless, a "stitch in space-time" can be illustrated diagrammatically. Charged particles are shown as dashed lines, where the dashes represent the discrete intervals of time in the particle's time line. Photon exchange is shown by the continuous grey lines. A single stitch such as that shown can only give a single value of distance and time for the second particle. Many stitches will be required to determine the properties of space-time. It is not necessary to assume that all these stitches require the immediate return of a photon, only that they use photon exchange and combine to give a consistent system of coordinates.

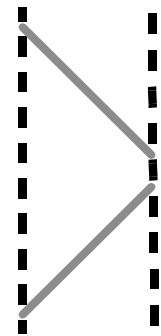

**Figure 6**

It is legitimate to imagine a system of such particles in the form of a diagram provided that it is understood that the space between the lines of the diagram has no theoretical or ontological meaning. The properties of space-time depend on internal relationships between dashes and nodes in the diagram, not on the geometry of a drawing. Thus in figure 7 the increased lapsed time for the return of a photon indicates that the particles are moving away from each other.

In systems of many particles such as we generally observe, photons are constantly exchanged, and macroscopic space-time can be construed as some kind of composition or average of the primitive space-time associated with photon exchange. Obviously we cannot empirically analyse an individual exchange of photons, since observation would disrupt it. But we can statistically analyse the effects of many such exchanges. Since the process of photon exchange is the same as we use in radar, the average behaviour of a system in which there are many such exchanges should obey the geometrical relationships found in the laws of Minkowsky space-time.

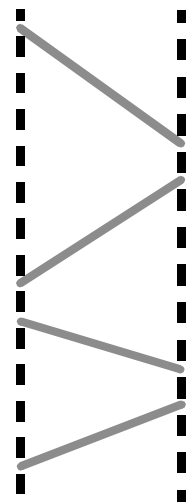

**Figure 7**

In many ways the current picture resembles the classical picture of point-like elementary particles, but now instead of particles at unknown co-ordinates we have particles in an unknown structure. We can partially determine the structure and can make statistical predictions about the behaviour of the structure, but since these predictions are not determined by unknown variables, classical probability theory does not apply. Intuitively we may hope to establish theoretical ground for the laws of quantum probability by the examination of properties abstracted from the structure. Since the macroscopic reference frame constitutes some form of average of the behaviour of individual particles and each individual particle is in part responsible for the generation of the macroscopic reference frame, individual particles do not in general have exact position and uncertainty in position becomes a feature of the description of elementary particles. Mathematical uncertainty is the subject of many valued logic [9]. The effect of many valued logic is that when an individual particle is viewed from a macroscopic reference frame it appears defocussed, or fuzzy. That is to say position is described by a many valued logical truth value. The argument that the logical truth function for measurement of position in a such a system obeys fundamental principles of quantum logic is given in [10].

Although it is not easy to see how to derive the four dimensional properties of Minkowsky space-time or spin from the notion of photon exchange between charged particles, it is possible to take the argument the other way. In [11] I have shown that empirically established properties of the measurement of space-time actually require an underlying substructure of matter in which leptons emit and absorb photons. Newtons laws and Maxwell's equations can then also be derived for the same structure. The derivation rests heavily on the uniqueness of the Dirac equation [12] and strongly suggests that the reason for a four dimensional universe with spin is that this is the smallest, and possibly the only, number of dimensions in which a solution exists.

It is possible to modify the treatment to take account of the time interval between the absorption and emission of photons implicit in reflection at the second charged particle. When this is done it is found that the geometrical relations are non-Euclidean and result in general relativity and the force of gravity. This is the subject of a paper in draft.